# Miras

**Lee Anne Willson**
**Massimo Marengo**
*Iowa State University, Department of Physics and Astronomy, Ames, IA 50011;*
*address email correspondence to lwillson@iastate.edu*

**Abstract** Mira variables share essential characteristics: High visual amplitude, periods of hundreds of days, red colors (spectral types M, S, and C), and the presence of emission lines at some phases. They are fundamental mode pulsators, with progenitor masses ranging from <1 to several solar masses. In this review, we summarize what is known from modeling and observational studies, including recent measurements from optical and IR interferometry, and studies involving large samples of stars particularly in the Magellanic Clouds. While we have a good idea of how these stars fit into the big picture of stellar evolution, many important details remain to be settled by a combination of more ambitious models and new observational techniques. Carrying on observations of bright Mira variables will be essential for interpreting observations of large numbers of fainter sources as well as for assessing the completeness and accuracy of the models.

## 1. Miras—fundamentals, mostly old news

Oxygen-rich Mira variables are red giants with visual amplitudes of 2.5 to > 7.5 magnitudes and relatively stable light curves; this corresponds to visual brightness changes from ×10 to ×1000. Their bolometric magnitude variation is about 1 magnitude, meaning the luminosity is varying only by a factor of 2–3. The reason that the visual brightness changes so much more over the cycle is that there are strong variations in the opacity of the atmosphere from minimum to maximum light, largely the result of variable amounts of TiO and other molecules.

There are fewer carbon-rich Miras, although plenty of variable carbon stars that are very similar to Miras in their bolometric variability. The difference in the atmospheric chemistry when C/O is > 1 vs. < 1 is part of the reason for this difference; the other is probably the sensitivity of the stellar radius to C/O, meaning the stars migrate across the zones of pulsational instability as C/O increases.

The pulsation of the Mira variables is driven by the same kind of process that drives the pulsation of Cepheids, and is a result of changes in the opacity and the dependence of pressure on temperature and density when hydrogen and helium are ionizing or recombining. In the Cepheids, the pulsation is rooted in the He+ <–> He++ zone, and in the Miras, in the H <–> H+ and He <–> He+ zones. Early modeling that showed this was done by Wood (1974; also Fox and Wood 1982) and by Ostlie (Ostlie and Cox 1986; Cox and Ostlie 1993).

The mode of pulsation is usually derived by taking the observed period, mass, and radius and comparing this with the periods derived from models with the same M and R. Thus, for red giants with M and R appropriate for Mira variables Ostlie and Cox found:

Fundamental mode
$P = 0.012d \, R^{1.86} \, M^{-0.73}$ or $\log P = -1.92 - 0.73 \log M + 1.86 \log R$

First overtone
$P = 0.04d \, R^{1.5} \, M^{-0.5}$ or $\log P = -1.40 + 1.5 \log R - 0.5 \log M$

These are both radial modes—that is, all the motion is in and out—and differ in that for the fundamental mode, all parts of the star move out or in together while for the first overtone, part of the star moves out while another part moves in. Thus the overtone mode has a shorter effective wavelength and if we assume the pulsation is like a sound wave, with the same sound speed, shorter wavelength → shorter period.

For the Miras, there are two difficulties with this approach. One problem is that we don't have direct measurements of the masses of the stars, and the other is that the radius measurements can be ambiguous or misleading. These difficulties were the cause of a long debate over the correct mode assignment for the Mira variables. In recent years the radius measurements have been improved and disambiguated via interferometry, and we have some new constraints on the current masses—see section 3. However, the debate was mostly settled before these results came in, based on modeling of shock waves in the atmosphere and a better understanding of the dynamics of the stellar atmospheres and winds.

The gist of the dynamical argument for the pulsation mode is this: When material in the atmosphere goes through one of the pulsation-induced outward-propagating shocks, it is given a kick and begins to travel outward. This trajectory is close to ballistic, that is, like a ball tossed upward. The stellar gravity acts on the material, bringing it back to its original position, or close to it, in time for the next shock to hit—just as hitting a ping-pong ball with a paddle can keep it in the air indefinitely if you always hit it at the right moment at the same place. For a purely ballistic motion, the infall velocity when the next shock comes through is equal to the outward motion after the previous shock, so the change in velocity is from +v to –v or 2v. This is the shock velocity amplitude, and can be deduced observationally from the Doppler shifts in spectral lines. The observed shock amplitudes are sufficiently large that they dictate a large gravity, inconsistent with the radii that give the observed periods with reasonable guesses about the masses of these stars (Hill and Willson 1979 and Willson and Hill 1979).

There are other ways to get at least a clue about the mode of pulsation. Typically, for fundamental mode pulsation, the motions are large enough to produce an asymmetric light curve—this is seen in Cepheids, RR Lyrae stars, and Miras. Overtone pulsation, in contrast, tends to produce more symmetric or sinusoidal variation and also smaller amplitude variability—this is definitely true when M and R are fixed, and still likely if P is held constant while M and/or R are varied.

In section 3 we will discuss at somewhat greater length the results of interferometric studies deriving the stellar radii at various phases of the pulsation cycle and some of what this has taught us about these stars. In section 2 we briefly review recent evolutionary modeling of relevance for Miras; in section 4, light curves and their shapes and secular changes; in section 5, the relation of these stars to RV Tauri stars and to planetary nebulae (if any), and in section 6 we discuss the recent waves of observational data from surveys and what such data can tell us when carefully analyzed.

The mass of a Mira variable is less than or equal to the mass of the star when it was on the main sequence. If we had a reliable formula for the mass loss as a function of mass, radius, luminosity, and so on, then we could integrate over the evolution and get a current mass—but we do not have such a formula (Bowen and Willson 1991; Willson 2000, 2009). If we had very good constraints on their radii we could get estimates for the masses from their pulsation periods; however, as noted above, radii are uncertain.

We do have a handle on the progenitor masses based on the distribution of the Miras in the galaxy—in simplest terms, we assign Miras to an appropriate population (old, young) based on their galactic orbits and on their metallicity. In our galaxy there is a correlation of age (older = lower mass) and metallicity in the sense that the shortest-period Miras have the lowest masses and metallicities. This fits with the age-metallicity correlation in general for field stars, and suggests the progenitors of the 200- to 250-day Miras were main sequence stars with masses near 1 solar mass while the 400- to 600-day Miras come from stars with progenitor masses ≥ 2 solar masses.

The assignment of a progenitor mass to a particular star based on its period can be risky, however, as these stars experience period changes during the course of shell flash cycling (see section 2). A star that normally sits at 500 days will spend up to 10% of its time between 200 and 300 days, for example. So at least the shorter-P population is expected to be contaminated with some higher-progenitor-mass interlopers.

Some recent attempts to derive masses for particular Miras based on interferometric observations have yielded masses close to the expected progenitor masses—something that we expect from the period-luminosity relation but that is nice to have confirmed by a more direct measure. Thus, for example, Lacour *et al.* (2009) find a mass of about 2 solar masses for χ Cyg, consistent with its relatively long period and large amplitude.

## 2. Modeling Miras with an emphasis on recent progress

There are three categories of models that are relevant for the study of Miras. The first is evolutionary models—following a star as it progressively uses up its nuclear fuels. The second is pulsation modeling, including testing very small perturbations for whether they will grow or damp out, and nonlinear modeling that seeks to determine the full-amplitude behavior of the pulsating star. The third is modeling of the atmosphere and wind, for the dual purposes of reproducing observed light curves, colors, and spectra and of determining the mass loss rates and velocities of the outflow for stars with a range of properties.

Miras are stars at the tip of the asymptotic giant branch; this much was known by the time LAW began to work on these stars in the 1970s. From evolutionary models we know that a star that begins with about 1 solar mass—a typical Mira progenitor star—will convert H to He in its core on the main sequence, taking about 10 Gyr to achieve this very slow "burning" of H. Then, the now predominantly He core will collapse into a degenerate state about the size of Earth and the star will expand to become a red giant. As to why it makes this transition, and fairly quickly, see Renzini and Ritossa (1994) and Iben (1993) for some authoritative arguments; a classic review of the evolution of low to intermediate mass stars is Iben (1967).

The source of energy for the red giant is the conversion of H to He in a shell around its degenerate core; thus the core mass gradually increases as it collects the garbage of the H-burning process. When the core of degenerate He reaches about half a solar mass, conversion of He to C and O begins by the triple-alpha process, where a very temporary He+He pair in the form of unstable $^{8}$Be is joined by a third He before it breaks up, and where the resulting C can add one more He to form O. Because the core is degenerate, it does not quickly readjust its structure when this new energy source turns on, and so, the process runs away—we call this the "Helium core flash." While one might expect this to explode the star, it does not, but after a brief disequilibrium the star settles into quiescent He burning as a horizontal branch star or clump giant, the extent to which it leaves the red giant track depending on the mass it has at this point.

When the He in the core has become C and O, the core again settles into a dense, degenerate state and the star once again evolves with increasing L up the "asymptotic giant branch" (AGB), so called because the track gradually converges to the earlier red giant track. Now, the conversion of H to He alternates with the conversion of He to C and O in a series of shell flashes or thermal pulses. (These pulses, which take perhaps 100,000 years to complete, are not to be confused with the pulsations that occur on a time scale of a few months or years.)

Some of the longer period Miras come from higher mass progenitors, 2 $M_{\odot}$ or more. Above about 2.8 solar masses the evolution is slightly different, with no He core flash. For a description of the evolution of these higher masses, see for example Iben (1975) or recent papers by Herwig (e.g. 2008) or Marigo *et al.* (2011, 2008). Grids of models are available online at various sites, including http://stev.oapd.inaf.it/cgi-bin/cmd (see Nasi *et al.* 2008), and an evolution code is available via the MESA project, http://mesa.sourceforge. net/, in two forms—a version for education and exploration and a version for serious research projects.

Red giants and AGB stars have degenerate cores surrounded by very deep convection zones reaching nearly all the way from the core to the surface, or from about 0.01 to 100 times the present solar radius $R_\odot$. Convection is very hard to model, and the behavior of the gas near the edges of the convection zone turns out to be critical for, for example, the contamination of the outer envelope and atmosphere with the products of the nuclear burning. One consequence of such contamination is a gradual increase in the ratio of carbon to oxygen, C/O. Carbon stars have C/O > 1 and S stars have C/O ≈ 1 while most stars, including M-type Miras, have C/O < 1. As C/O changes, so does the internal opacity and thus the radius of the star for a given L and M (Marigo and Girardi 2007). Also, because C and O form a very stable molecule, CO, the chemistry of the surface layers, atmosphere, and wind changes dramatically as C/O passes 1, and the kinds of grains that form and assist in the mass loss process also change.

Evolutionary modeling in recent years has focused on the convection process, particularly on the effects of "convective overshoot" mixing material beyond the boundaries of what are otherwise the domains of convective instability. There have also been advances in modeling the variations that occur during He shell flashes. Very little has been done to study the pulsation in more detail—the linear analyses done in the late 1970s and 1980s give useful results, and models with full-up non-linear non-adiabatic pulsation in a fully convective envelope are just out of reach of existing codes and machines, although we expect advances in this area in the next decade.

## 3. Interferometry and other methods for probing the near-star environment

Interferometry is a means of seeing detail not possible with a single aperture or telescope. Long baseline interferometry, using two or more telescopes, provides the finest spatial resolution. Aperture masking and segment tilting interferometry, that involves dividing the mirror of a very large telescope into smaller sub-apertures (using a mask, for example), has also been used to study Miras (Haniff *et al.* 1992; Woodruff *et al.* 2008, 2009). Earlier papers also cite "speckle interferometry" which is not really interferometry at all but rather a method for getting around the effects of seeing by adding short exposure sub-images together (Labeyrie 1970; Bonneau and Labeyrie 1973).

To resolve the diameters of Miras requires milli-arcsec resolution even for the nearest stars (χ Cyg, R Leo, and Mira). Resolved diameters yield more than just a number; watching the variation of the diameter over the cycle allows for determination of $\Delta R/R$ which, with observed velocities, can yield R and thus distance as well as a constraint on pulsation models. Another constraint on models comes from the variation of the apparent angular diameter with the wavelength of the observations, telling us how the atmospheric opacity varies with wavelength and revealing molecule-rich or dust-rich layers in the atmosphere.

Starting from the measurement of o Ceti's diameter by Pease in 1931 using an optical Michelson interferometer, the main goal of these studies was to resolve the controversy on Miras' pulsation modes by measuring their radius. While early results (Tuthill *et al.* 1994) measured overly large radii, suggesting that Mira variables should be first overtone pulsators, subsequent analysis (Menesson *et al.* 2002; Perrin *et al.* 2004) found that these diameters were actually biased by the presence of molecular layers (mainly CO and $H_2O$) a few stellar radii above the photosphere. This breakthrough was enabled by the introduction of a new generation of instruments at world's largest interferometers (VLTI in the southern hemisphere and PTI/CHARA in the north), capable of simultaneously measuring the angular diameter at several broad and narrow bands. These observations can be directly compared with time-dependent hydrodynamic simulations (see, for example, Hillen *et al.* 2012), providing precious observational constraints to stellar models and pulsation theory (Le-Bouquin *et al.* 2009; Martí-Vidal *et al.* 2011) and allowing the exploration of non-equilibrium chemistry in the stellar atmosphere (Paladini *et al.* 2009).

Interferometry with more than two elements also yields information about departures from spherical symmetry in the system. Ragland *et al.* (2006) combined the light from the three apertures of the IOTA interferometer to study the asymmetry of M- and C-type giants with different pulsation properties. Non-zero closure phase (a signature of departure from spherical symmetry) was found in 30% of the targets, or essentially all that were reliably resolved. This asymmetry did not depend on the chemistry of the atmosphere, but was much more common among Miras than other types of long period variables. These asymmetries were located in close proximity of the stellar surface (1.5–2 stellar radii) suggesting an origin either related to the presence of large convective cells in the stellar photosphere, discrete dust clouds formation, or the interaction with a companion. Recent results using aperture synthesis techniques borrowed from radio-interferometry have allowed the reconstruction of true images for a few late-type pulsating variables (for example, χ Cyg in Lacour *et al.* 2009, or RR Aql in Karovicova *et al.* 2011). These images have provided direct evidence for the presence of hot cells in the stellar atmosphere and, combined with radial velocity measurements, have allowed the mapping of speed, density, and position of diverse molecular species in pulsation-driven atmospheres.

Asymmetric dust shells have also been detected around several targets, using either the thermal-infrared long baseline interferometry, or speckle, aperture masking, or segment-tilting techniques on large-aperture telescopes. The recent detection of large, transparent grains as close as 1.5 stellar radii around a number of mass-losing cool giants (including the carbon Mira R Leo, Norris *et al.* 2012) shows the potential of these techniques for probing the dust formation processes that are key for understanding mass loss in Mira variables. While the current angular resolution is not sufficient to measure the motion of these dust layers during the pulsation cycle, the ultimate goal of these observations is to fully characterize the stratification, geometry, and kinematics of the extended Miras' atmosphere, in order to finally understand how mass loss processes are connected to the stellar pulsations, and the root causes for the asymmetries and inhomogeneity observed in their circumstellar environments (Wittkowski *et al.* 2012).

## 4. Light curves—shapes and secular changes

In addition to the dominant signature of pulsation, Mira variable light curves show an interesting variety of features. The ratio of rise time to cycle length, called *f* by Campbell (1955), varies over a range from about 0.5 down to 0.1 or 0.2 among some of the longest-period Miras. Some of the stars show bumps on their rising or falling branches, and these bumps may come and go over the course of a number of cycles—see, for example, the long-term light curve for χ Cyg. Some of the stars show long-term modulations of their periods or their amplitudes or their mean magnitudes, and these variations are not at all well understood. Some show steady decreases or increases in the periods, often with correlated amplitude changes. Some stars behave like regular Miras for some decades, then fairly abruptly switch to smaller amplitude and/or shorter period variation, causing them to be reclassified as SR stars; it is not clear whether, if we waited long enough, this would happen to a large or a small fraction of our Mira stars.

Templeton *et al.* (2005) looked at a century of AAVSO data for Miras and found that about 10% show significant long-term variability of their periods. The specific variations are, however, all over the map, from apparently sinusoidal (where we need another century to be sure) to apparently steady decrease or increase to abrupt changes from constant to steeply decreasing. While about 1% of the stars are expected to be in a rapidly varying phase of the shell flash cycle, and about 1% of the light-curves show period variations consistent with this explanation, the other 9% of Miras show period variations that are unexplained at present.

## 5. Relation of Miras to other classes of variables and to planetary nebulae (not)

SR variables differ from Miras in that they have smaller amplitudes, less stable light curves, or are not red giants. Classically, smaller amplitude ones are SRa, less stable ones are SRb, and supergiants are SRc. However, there is substantial evidence that these distinctions are not the best for separating physically different objects, as some SRa appear to be very much like Miras apart from their visual amplitudes; some Miras start behaving erratically and are reclassified as SRb; and the sequence of pulsating AGB stars extends perhaps as high as seven or eight solar masses, making the distinction with SRc also a bit unclear. Some excellent discussions of these classes of objects may be found in the papers by Hron and Kerschbaum (Kerschbaum and Hron 1996; Kerschbaum *et al.* 1996; Hron *et al.* 1997; Lebzelter *et al.* 1995).

The RV Tauri stars are thought to be a post-Mira, possibly pre-planetary-nebula phase of evolution. The light curves have some characteristics in common with some Miras, including double maxima and long term modulations. Not all Miras become RV Tauri stars, only those from the low-mass, low-luminosity, low-metallicity end of the Mira distribution. For a discussion of RV Tauri stars and their relation to Mira variables see Willson and Templeton (2009).

While it is often stated that the same stars that go through a Mira stage later become central stars of planetary nebulae on their way to becoming white dwarf stars, recent work suggests that only a fraction of Miras can become the central stars of PNe and quite possibly none of the classical Miras will do so; it now appears likely some form of binarity is required for all but perhaps the highest mass progenitors of PNe. This conclusion is reached after recognizing that there are only about 15% as many PNe produced per year as there are stars leaving the AGB per year, so at most 15% of the Miras can become the central stars of PNe. Also, the time scale for evolution across the HR diagram after the AGB is too slow for most of the stars, being perhaps long enough for just the higher masses; finally, the fact that most PNe are not spherically symmetric hints at more than a single-star origin for the PNe. See, for example, papers presented at IAU Symposium 283, Planetary Nebulae, an Eye to the Future (held in 2011, proceedings soon to appear).

## 6. Using photometric surveys to characterize the Mira and carbon star components of populations

One of the most powerful tools to study variable stars populations in our and other galaxies is represented by unbiased multi-epochs photometric surveys at optical and infrared wavelengths. If the distance to the stars is known (for example, for stars in the same galaxy or cluster), it is then possible to plot the absolute magnitude of the stars as a function of log period. Pulsators with different mode and amplitude tend to be distributed on separate sequences, corresponding to different period-luminosity relations. This approach has been famously pioneered by Peter Wood, using the MACHO database for Large Magellanic Cloud (LMC) variables. Miras appear to be distributed at the bright end of a linear sequence (the "C" sequence as defined in Wood 2000) populated by fundamental mode pulsators, with semiregular (SR) variables at the lower, fainter end of the sequence. Recent results from the OGLE-III survey (Soszyński *et al.* 2009, 2011) have confirmed these results and extended them to the Small Magellanic Cloud (SMC). This analysis is less successful for galactic Miras, due to the uncertainties in their distance stemming from the poor quality of their Hipparcos parallax. The Gaia mission, by collecting accurate parallaxes of Mira variables across the whole disk of our Galaxy, will be an invaluable resource towards the calibration of a precise period-luminosity relation, on par to what is currently available for Cepheids (Whitelock 2012). Given their higher intrinsic luminosity than other standard candles (Cepheids and RR Lyrae), Miras can be adopted as probes for galactic structure, and precise distance indicators for the farthest galaxies in the local group.

Optical and near-IR surveys offer only a limited capability in discriminating between M-type and carbon Miras. Thermal infrared photometry, on the other end, allows probing the wavelength range where dust features of mass-losing Miras are stronger, providing an excellent diagnostic for their extended atmosphere. Period-luminosity diagrams made for the LMC using Spitzer Space Telescope data show similar sequences as Wood (2000), but an increased level of separation between carbon- and oxygen-rich Miras (Riebel *et al.* 2010). Color-color and color-magnitude diagrams for both the LMC and the SMC in Spitzer bands (Blum 2006; Boyer 2011) provide an even better separation, and allow studying the statistics of carbon and M-type Miras at different metallicity. The main result of these surveys is the confirmation that a low metallicity environment favors the formation of carbon Miras, due to the smaller number of third dredge-up events that are required to push the C/O ratio above unity. These surveys have also highlighted a population of long period variables with very large infrared excess, usually referred to as "extreme AGB" stars. These objects, presumably associated with very large mass loss rates, in the LMC and SMC are generally characterized by a C-rich dust chemistry. While some of these sources lie at the top of the fundamental mode period luminosity sequence, most of them are overtone pulsators (Riebel *et al.* 2010). This result may, however, be the consequence of observational biases preventing the redder sources to be detected in optical and near-IR surveys like MACHO and OGLE-III.

## 7. Final comments

As Figure 1 shows, research on Miras has waxed and waned several times during the century of the AAVSO. Generally, new observing capabilities or much improved modeling codes lead to increased activity, followed by a decline in action until the next advance. Studies of individual stars and systems are fed by spectroscopy, interferometry, and detailed atmospheric modeling, while studies of the evolution of the stars and the consequences of mass loss associated with the Mira stage are being advanced by evolutionary modeling, infrared observations, and the statistical analysis of large populations of variable stars.

The AAVSO has played and continues to play a central role in the study of Miras. Most of the papers in Figure 1 are observational; theory papers are a minority. Indeed, in 1972 when LAW's first refereed paper appeared, apparently one eminent astronomer commented "There's a theorist working on Miras. He must be out of his mind!" Nearly every observational study of galactic Miras includes reference to phases or light curves based on data collected by AAVSO or its sister organizations around the world. While the massive studies of Magellanic cloud Miras based on the dark-matter-search byproducts and the huge IR surveys turning up many more of these highly evolved stars described in section 6 mostly do not directly refer to AAVSO data, interpreting the resulting period-luminosity plots and understanding how the Miras and their SR cousins are related will again depend on long-term observations and studies of bright, relatively nearby stars. Future large surveys, including the LSST (Large Synoptic Survey Telescope), will focus on very faint stars and have a duration of observations ranging from a few years to a couple of decades; AAVSO already has a century of data. For stars with periods around a year, a decade is a very short window of observation.

At the present time the observational technologies are expanding our ability to probe the circumstellar environment, to study large samples of stars in a systematic way, and to model the essential physical processes in the interiors, atmospheres, and winds. However, in all of these areas much remains to be done as we are still limited by the capabilities of our computers, detectors, and telescopes.

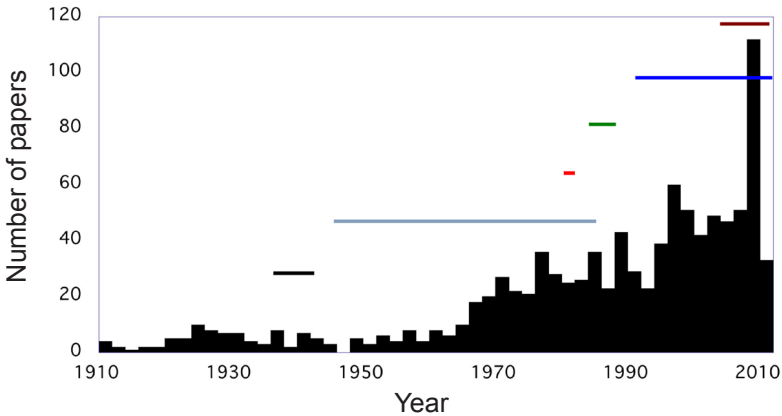

Figure 1. 1,000 refereed papers pulled up by ADS on a search for the words "Mira," "Long period variable," or "AGB star" in the title, from 1910 to 2010, plotted as a function of time. Some events that may have influenced the graph include (horizontal bars) WW II 1939–1945, the Palomar era 1949–1992, IRAS 1983, Hipparcos 1989–1993, the dark-matter searches MACHO, OGLE, and so on (starting 1993), and Spitzer (from 2003). IRAS and Spitzer, being infrared missions, definitely boosted interest in these stars. A big surge also came about three years after the beginning of the MACHO, OGLE, and similar projects.